\documentclass[prb,twocolumn,showpacs,nofootinbib,preprintnumbers,amsmath,amssymb,aps,longbibliography,floats]{revtex4-2}
\usepackage{graphicx,subfigure,epsfig}
\usepackage{dcolumn}
\usepackage{amssymb}
\usepackage{times}
\usepackage{amsmath}
\usepackage{amsthm}
\usepackage{amsfonts}
\usepackage{mathrsfs}
\usepackage{setspace}
\usepackage{latexsym}
\usepackage{bbm}
\usepackage{float}
\usepackage{flafter}
\usepackage{bm}
\usepackage{epstopdf}
\usepackage{color}
\usepackage{multirow}
\usepackage{physics} 
\usepackage[export]{adjustbox}
\DeclareMathOperator{\sgn}{sgn}
\usepackage{braket}

\newtheorem{theorem}{Theorem}[section]
\newtheorem{lemma}[theorem]{Lemma}

\newcommand{\bk}{{\bf k}}
\newcommand{\bq}{{\bf q}}
\newcommand{\bp}{{\bf p}}
\newcommand{\bv}{{\bf v}}
\newcommand{\bs}{{\bf s}}
\newcommand{\br}{{\bf r}}

\usepackage{xr}
\usepackage[pdftex,breaklinks,bookmarks=false,colorlinks,linkcolor=blue,citecolor=blue,urlcolor=blue,hypertexnames=false]{hyperref}

\usepackage{orcidlink}

\begin{document}

\title{Three-point density correlations in a weakly interacting 2D Fermi liquid}

\author{C. L. Kane \orcidlink{0000-0002-9551-2177}}
\affiliation{Department of Physics and Astronomy, University of Pennsylvania, Philadelphia PA 19104}

\date{\today}

\begin{abstract}
We study the three-point equal-time correlations of the density in a weakly interacting spin 1/2 Fermi gas and present two new results.   First, we compute the three-point correlation $s_3^\rho(\bq_1,\bq_2)$ for the total density $\rho = \rho_\uparrow + \rho_\downarrow$ exactly as a function of $\bq_1$ and $\bq_2$ to first order in a dimensionless interaction parameter ${\cal I}$.   This generalizes a previous result that related the singularity in $s_3^\rho(\bq_1,\bq_2)$ to the Landau Fermi liquid parameters $F_0^s$ and $F_0^a$ that applied in a certain long-wavelength collinear limit of $\bq_1$ and $\bq_2$.    Second, we compute the leading order ${\cal O}({\cal I}^3)$ interaction correction to the same-spin three-point correlation function $s_3^\uparrow(\bq_1,\bq_2)$ in the long-wavelength collinear limit.    These results are directly relevant to current experiments on atomic Fermi gases using quantum gas microscopy.
\end{abstract}

\maketitle

\section{Introduction}

Correlation and response functions provide a powerful tool for probing the structure of quantum phases of matter.   Gapless quantum states exhibit long range correlations that reflect universal features of the underlying phase of matter.   Two-point correlations, which are related to linear response coefficients, have long been studied in this connection\cite{agd,nozieres1999theory,mahan2000many,Giuliani_Vignale_2005}.   Three- and higher-point correlations, which are relevant to interaction effects as well as non-linear response, have also attracted interest\cite{brovman1973singularities,brovman1974general,Metzner1998PRB,feldman1998evaluation,Theumann1984,Metlitsky2010,Metzner2011}. 

In recent work, we showed that the multi-point correlations of a non-interacting Fermi gas reflect the topology of the Fermi sea\cite{TamKane2022a,TamKane2023b}.   In two dimensions, the equal-time three-point correlation function of the charge density, $s_3(\bq_1,\bq_2)$, 
is given by the `topological formula' $\chi_F |\bq_1\times \bq_2|/(2\pi)^2$, provided $\bq_1$ and $\bq_2$ are small enough to be in the `topological regime'.   
The coefficient $\chi_F$ is the Euler characteristic, which is an integer quantized topological invariant that characterizes the Fermi sea in momentum space.
The $|\bq_1\times\bq_2|$ in the topological formula defines a unique type of singularity, which in real space corresponds to correlations that are significant when the three real space positions $\br_{1,2,3}$ are far from each other, but are oriented along the same straight line.   These long-ranged collinear correlations are a unique signature of the Fermi surface, which features low energy excitations that propagate ballistically, and they play an important role in other proposed consequences of Fermi sea topology in Fermi gases\cite{Kane2022a,TamKane2022b,Yang2022,Zhang2023,TamKane2023a,yang2025quantized}.

This result is modified when the fermions interact.  While in general the interaction corrections depend on the details of the Hamiltonian as well as of the interactions, it was argued\cite{TamKane2026} that in a Fermi liquid, the three-point correlations retain a universal singular behavior that is revealed in a long-wavelength collinear (LWC) limit.   The singularity in $s_3(\bq_1,\bq_2)$ in the LWC limit depends only on the Landau parameters that characterize the low energy Fermi liquid fixed point\cite{agd,landau1957theory,landau1959theory}.   This result complements earlier work characterizing singularities in Fermi liquids\cite{Chubukov2003}, and it is reminiscent of the well understood behavior of an interacting one-dimensional Fermi gas, where the long-wavelength density correlations are determined by the Luttinger parameter $K$, which characterizes the low energy Luttinger liquid fixed point\cite{haldane1981luttinger,Giamarchi2004, gogolin2004bosonization}.    

Multi-point density correlations have recently been measured in a degenerate Fermi gas of $^6$Li atoms using quantum gas microscopy\cite{daix2025probing}.  The measured three- and four-point correlations agree quantitatively with the free Fermi gas predictions, both in the topological regime and outside the topological regime.    Quantum gas microscopy experiments offer a powerful method for studying the effects of interactions because the interactions can be controlled by tuning a magnetic field in the vicinity of a Feshbach resonance\cite{bakr_review,Cheuk2015,Edge2015,Haller2015,Omran2015,Parsons2015,Cheuk2016a,Cheuk2016b,deJongh2025,daix2025observing}.   The interactions are essentially a contact interaction, in which only fermions with opposite spin interact.   The interaction can be characterized by a dimensionless parameter ${\cal I} = -1/\log k_F a$, which is related to the scattering length $a$ for particles at the Fermi energy.   The experiments measured the three-point correlations between particles with the same spin $s_3^\uparrow(\bq_1,\bq_2)$ in the presence of moderately strong attractive interactions, $-0.5 < {\cal I}<0$ which are in a regime  where the critical temperature for BCS pairing is well below the scale set by temperature and finite size.    Surprisingly, the agreement with the non-interacting Fermi gas prediction was excellent even for interactions as large as ${\cal I} = -0.5$.

The Fermi liquid theory of Ref. \cite{TamKane2026} provided some explanation for this apparent insensitivity to interactions.    It was found that in the LWC limit, the three-point same-spin correlations $s_3^\uparrow(\bq_1,\bq_2)$ and three-point correlations of the total density $s_3^\rho(\bq_1,\bq_2)$ have a different dependence on the Landau parameters $F_0^s$ and $F_0^a$.   Using the known perturbative expansion of the Landau parameters to second order in ${\cal I}$, it was observed that while $s_3^\rho$ exhibits a correction at linear order in ${\cal I}$, the first and second order terms of $s_3^\uparrow$ both vanish.   This suggests that measuring the three-point correlations of the total density provides a more sensitive probe of the effects of interactions than the same-spin correlations, and it poses the question:
 what is the leading order correction to the same-spin correlations?   

In this paper we present two new results that are motivated by these findings.    (i)  We determine the leading ${\cal O}({\cal I}^3)$ correction to $s_3^\uparrow(\bq_1,\bq_2)$ in the LWC limit by computing the necessary perturbative corrections to the Landau parameters.   It turns out that the new information that is required is the coefficient of the ${\cal O}({\cal I}^3)$ term of the Landau parameter $F_0^{\uparrow\uparrow}$, which characterizes the interaction between quasiparticles with like spins.   We will show that the ${\cal O}({\cal I}^3)$ term in $F_0^{\uparrow\uparrow}$ is in fact equal to zero.    This then leads to a non-zero prediction for the ${\cal O}({\cal I}^3)$ correction to $s_3^\uparrow$.

(ii)  We consider in more detail the effects of interactions on the total density correlations.   While the Fermi liquid theory presented in Ref. \cite{TamKane2026} is valid in the LWC limit, the experiments probe a larger parameter regime in which the LWC limit does not apply.   Here, we compute the total density correlation function exactly to first order in ${\cal I}$.   This new result reduces to the first order expansion of the Fermi liquid theory in the LWC limit, but it also applies to more general values of $\bq_1$ and $\bq_2$, allowing for detailed comparison with experiment as well as providing new insight into the regime of validity of the LWC limit.   

The paper is organized as follows.   In Section \ref{Section 2} we specify the model, explain our notation and summarize previously known results.   In Section \ref{Section 3} we present the first order perturbative  correction to $s_3^\rho(\bq_1,\bq_2)$, and in Section \ref{Section 4} we present the ${\cal O}({\cal I}^3)$ correction to $s_3^\uparrow(\bq_1,\bq_2)$.  Some of the calculations were involved, so  further details will be provided in the appendices.
    
\section{Preliminaries}
\label{Section 2}

\subsection{Model and Notation}

We consider an interacting spin $1/2$ Fermi gas described by the Hamiltonian
\begin{equation}\begin{split}
    {\cal H} = \sum_{\bk} &E_\bk(c_{\bk,\uparrow}^\dagger c_{\bk,\uparrow}+c_{\bk,\downarrow}^\dagger c_{\bk,\downarrow})\\
   & + u \sum_{\bk,\bk',\bq} c^\dagger_{\bk+\bq,\uparrow}c_{\bk,\uparrow} c^\dagger_{\bk'-\bq,\downarrow}c_{\bk',\downarrow},
\end{split}\end{equation}
with $E_\bk = |\bk|^2/2m$.  Here and in the following we set $\hbar=1$.  The second term describes a contact interaction that is a $\delta$-function in real space and only couples opposite spins.

The central quantities of interest are the equal-time three-point correlation function of the total density, and the same-spin three-point correlation function.   The total density correlator is defined by 
\begin{equation}
 s_3^\rho(\bq_1,\bq_2) \equiv \int \frac{d^2\bq_3}{(2\pi)^2} \langle \rho(\bq_1)\rho(\bq_2) \rho(\bq_3)\rangle_c
\end{equation} 
where the total density is $\rho(\bq) = \rho_\uparrow(\bq) + \rho_\downarrow(\bq)$ with $\rho_\sigma(\bq) = \sum_\bk c^\dagger_{\bk+\bq,\sigma}c_{\bk,\sigma}$ for $\sigma = \uparrow,\downarrow$.  The subscript $c$ denotes a connected correlation function, which is proportional to $\delta(\bq_1+\bq_2+\bq_3)$.   The same-spin correlator is  
\begin{equation}
s^\uparrow_3(\bq_1,\bq_2) \equiv \int\frac{d^2\bq_3}{(2\pi)^2} \langle \rho_\uparrow(\bq_1)\rho_\uparrow(\bq_2) \rho_\uparrow(\bq_3)\rangle_c
\end{equation}
Our goal is to understand the behavior of these quantities in a perturbative expansion in powers of the interaction.

It is well known that a perturbative expansion of the two dimensional Fermi gas in powers of $u$ is divergent in the ultraviolet for an infinitely short-ranged interaction\cite{Randeria1992}.   This divergence can be repaired by considering a renormalized theory that is expressed in terms of the physical scattering amplitudes for particles near the Fermi energy\cite{agd, AbrikosovKhalatnikov1957}, which are not divergent.   This is accomplished by introducing an ultraviolet momentum cutoff $k_\Lambda$ (or equivalently energy cutoff $E_\Lambda = k^2_\Lambda/2m$), and considering the two-particle scattering problem.
For the contact interaction in two dimensions with a constant density of states (per spin, per unit area) $N_0 = m/(2\pi)$, the $T$-matrix for two particles with momentum $\bk$ and $-\bk$ and total energy $2E$ is evaluated by summing the Born series for the retarded two-particle propagator, leading to
\begin{equation}
    T(E) = \frac{u}{1+\frac{1}{2} N_0 u \log (-\frac{E_\Lambda}{E+i\epsilon})} .
\end{equation}
For particles at the Fermi energy $E_F = k_F^2/2m$, we introduce the dimensionless interaction
\begin{equation}
    {\cal I} = N_0 {\rm Re}[ T(E_F)] \equiv - \frac{1}{\log k_F a},
\end{equation}
where the scattering length is $a = k_\Lambda^{-1} e^{-1/(N_0 u)}$ and we have assumed interactions are weak, $N_0 |u| \ll 1$, so that the imaginary part of the denominator is negligible.
In this case, for repulsive interactions ($u>0$) we have $k_F a \ll 1$.  For attractive interactions ($u<0$) $k_F a \gg 1$, reflecting the existence of a weakly bound two-particle bound state of size $a \gg k_F^{-1}$.

${\cal I}$ can be expanded perturbatively in powers of $u$, 
\begin{equation}
    {\cal I} = N_0 u - (N_0 u)^2 \log \frac{k_\Lambda}{k_F} + {\cal O}(u^3).
    \label{i to u}
\end{equation}
It can be seen that for $n\ge 2$ the ${\cal O}(u^n$) terms in the expansion exhibit a logarithmic divergence when $k_\Lambda \rightarrow\infty$.   
We will see that the expansions of $s_3^\rho$ and $s_3^\uparrow$ in powers of $u$ exhibit similar logarithmic divergences.   However by inverting (\ref{i to u}),
\begin{equation}
    N_0 u = {\cal I} + {\cal I}^2 \log \frac{k_\Lambda}{k_F} + {\cal O}({\cal I}^3),
    \label{u to i}
\end{equation}
the expansion of $s_3^\rho(\bq_1,\bq_2)$ and $s_3^\uparrow(\bq_1,\bq_2)$ in powers of ${\cal I}$ is finite.   This reflects the invariance of physical quantities under the renormalization group, where the dimensionless interaction is renormalized when the cutoff is rescaled according to $d{\cal I}/d \log k_\Lambda = - {\cal I}^2$\cite{Polchinski1992,Shankar1994}.   The strategy is therefore to compute the expansion of the three-point density correlation function along with the Landau parameters in powers of $u$, and then to re-express the results as an expansion in  powers of ${\cal I}$.    It is expected that the divergent $\log k_\Lambda/k_F$ terms will cancel at each order in ${\cal I}$.    For the first order expansion of $s_3^\rho(\bq_1,\bq_2)$ studied in Section \ref{Section 3}, this renormalization is unimportant, and the first term in (\ref{u to i}) suffices.   For the third order calculation of $s_3^\uparrow(\bq_1,\bq_2)$ in Section \ref{Section 4}, the second order correction in (\ref{u to i}) is required.

\subsection{Prior Results}

For non-interacting fermions with $u=0$, $s_3^\rho = 2 s_3^\uparrow \equiv 2 s_3^{(0)}(\bq_1,\bq_2)$  has been computed exactly\cite{TamKane2022a,TamKane2026}  
\begin{equation}
    s_3^{(0)} =\int \frac{d^2\bk}{(2\pi)^2} (1-f_{\bk+\bq_1}-f_{\bk+\bq_2})f_\bk \bar f_{\bk + \bq_1+\bq_2},
    \label{s30}
\end{equation}
where $f_\bk$ is the zero temperature Fermi occupation factor and $\bar f_\bk = 1-f_\bk$.
For the circular Fermi surface of radius $k_F$, when the `topological condition'
\begin{equation}
|\bq_1||\bq_2|\bq_3|\le 2 k_F |\bq_1\times \bq_2|
\label{topological condition}
\end{equation}
is satisfied (with $\bq_3=-\bq_1-\bq_2$), $s_3^{(0)}$ is given by the topological formula with $\chi_F=1$:
\begin{equation}
    s_3^{(0)}(\bq_1,\bq_2) = \frac{|\bq_1\times\bq_2|}{(2\pi)^2}.
    \label{top formula}
\end{equation}
For $|\bq_1||\bq_2|\bq_3|\ge 2 k_F |\bq_1\times \bq_2|$, the integral in (\ref{s30}) was evaluated in Ref. \cite{daix2025probing}, leading to
\begin{equation}
    s_3^{(0)}(\bq_1,\bq_2) = \frac{k_F^2}{4\pi}\left[1  + \sum_{a<b<c=1}^3 \sigma_{ab}G(\frac{|\bq_c|}{2k_F})\right],
    \label{i=0 formula}
\end{equation}
where $\sigma_{ab}={\rm sgn}(\bq_a\cdot\bq_b)$ and
\begin{equation}
    G(X) = \frac{2}{\pi}(\cos^{-1}X - X\sqrt{1-X^2})\theta(1-X).
\end{equation}

For an interacting Fermi liquid, it was argued that the three-point correlation function contains a universal singularity that is insensitive to irrelevant perturbations, and depends only on the Landau parameters that characterize the Fermi liquid fixed point\cite{TamKane2026}.    This singularity is isolated in the long-wavelength collinear (LWC) limit, in which $\bq_1$, $\bq_2$ and $\bq_3$ are small and nearly parallel (or anti-parallel), so that
\begin{equation}
|q_\perp| \ll |\bq_1|, |\bq_2|, |\bq_3| \ll k_F, 
\end{equation}
where $|q_\perp|$ is the maximum of $|\bq_a\times\bq_b|/|\bq_a|$ for $a\ne b = 1,2,3$.
In this limit, the three-point correlation of the total density can be expressed as
\begin{equation}
    s_3^\rho(\bq_1,\bq_2) = \frac{|\bq_1\times \bq_2|}{(2\pi)^2}\frac{2}{(1+F_0^s)^3},
\end{equation}
while the same-spin correlation is
\begin{equation}
    s_3^\uparrow(\bq_1,\bq_2) = \frac{|\bq_1 \times\bq_2|}{(2\pi)^2}\frac{(1+F_0^a)^2+3(1+F_0^s)^2}{4(1+F_0^s)^3 (1+F_0^a)^2},
    \label{s3up}
\end{equation}
where $F_0^{s,a}$ are the isotropic Landau parameter symmetric and antisymmetric in the spin.

The Landau parameters for the two dimensional Fermi liquid have been computed perturbatively in Ref. \cite{Randeria1992} to ${\cal O}({\cal I}^2)$.  It was found\cite{tarik} that
\begin{equation}\begin{split}
  &  F_0^s = {\cal I} + {\cal I}^2 (2- \log 2) + {\cal O}({\cal I}^3) ,\\
  &  F_0^a = -{\cal I}+ {\cal I}^2 \log 2 + {\cal O}({\cal I}^3).
  \label{f0s,f0a}
  \end{split}\end{equation}
This led to predictions for the three-point correlation functions in the LWC limit\cite{TamKane2026}:
\begin{align}
s_3^\rho &= 2 \frac{|\bq_1\times\bq_2|}{(2\pi)^2}\left[1 - 3{\cal I} + 3{\cal I}^2 \log 2 + {\cal O}({\cal I}^3)\right],\label{s3rho old}\\
s_3^\uparrow &= \frac{|\bq_1\times\bq_2|}{(2\pi)^2}\left[1 +  {\cal O}({\cal I}^3) \right].  
\end{align}

It is notable that for weak interactions, $s_3^\rho$ exhibits an interaction correction that is linear in ${\cal I}$, while the ${\cal O}({\cal I})$ and ${\cal O}({\cal I}^2)$ corrections to $s_3^\uparrow$ vanish.   The vanishing of $s_3^\uparrow$ at linear order follows simply for a contact interaction, which at linear order only couples opposite spins.   The vanishing at second order is less trivial, but some insight can be gained\cite{thankreferee} by expressing (\ref{s3up}) in terms of the scattering amplitudes $A_0^{s,a} = F_0^{s,a}/(1+F_0^{s,a})$, which are of order ${\cal I}$.   With this substitution, we obtain
\begin{equation}\begin{split}
    s_3^\uparrow = \frac{|\bq_1 \times\bq_2|}{(2\pi)^2}&\Big[1-\frac{3}{2}(A_0^s+A_0^a) + \frac{3}{4}(A_0^s+A_0^a)^2 \\
    & - \frac{1}{4}\left( (A_0^s)^3 + 3 A_0^s(A_0^a)^2\right)\Big].
    \end{split}\end{equation}
Thus, it can be seen that to order ${\cal I}^2$, the correction to $s_3^\uparrow$ is determined by $A_0^{\uparrow\uparrow} = (A_0^s + A_0^a)/2$.   It can be checked that for the contact interaction $A_0^{\uparrow\uparrow}$ vanishes to ${\cal O}({\cal I}^2)$.   This is related to a general constraint on the amplitude for forward scattering due to the Pauli principle,  $A^{\uparrow\uparrow}(\theta = 0) = 0$.  This leads to a constraint on the scattering amplitudes summed over angular momentum channels: $\sum_l A_l^{\uparrow\uparrow} = 0$ \cite{Mermin1967}.   To ${\cal O}({\cal I}^2)$ for the contact interaction this constraint is obeyed independently for each angular momentum channel.

The goal for the remainder of the paper will be (i) to generalize the linear in ${\cal I}$ correction to $s_3^\rho$ beyond the LWC limit and (ii) to determine the leading  ${\cal O}({\cal I}^3)$ correction to $s_3^\uparrow$ in the LWC limit by determining the necessary ${\cal O}({\cal I}^3)$ corrections to the Landau parameters.

\section{First Order Correction for Total Density Correlations}

\label{Section 3}

\begin{figure}
    \centering
    \includegraphics[width=.9\linewidth]{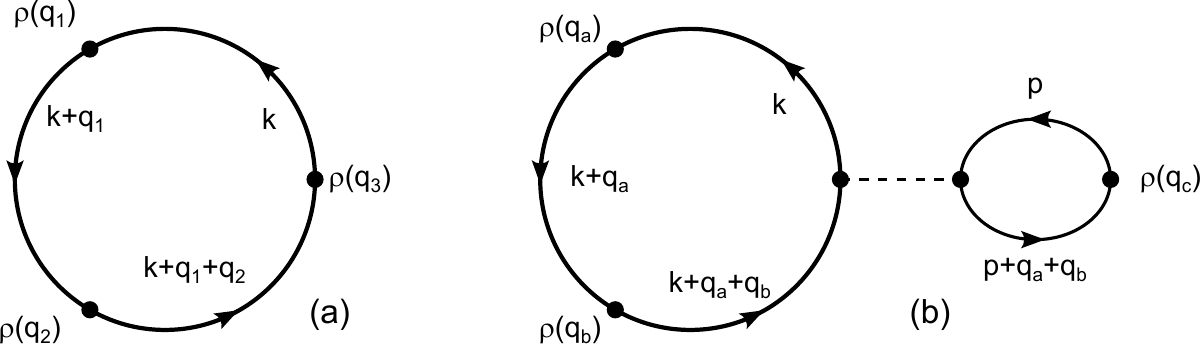}
    \caption{Feynman diagrams for $s_3^\rho$ to (a) zeroth order and (b) first order in the interaction $u$.}
    \label{1st order diagrams}
\end{figure}

In this section we compute the exact first order correction to $s_3^\rho(\bq_1,\bq_2)$, which generalizes the first order term in (\ref{s3rho old}) beyond the LWC limit.  The expansion may be written,
\begin{equation}
    s_3(\bq_1,\bq_2) = 2 \left[s_3^{(0)}(\bq_1,\bq_2) + {\cal I} s_3^{(1)}(\bq_1,\bq_2)\right] + {\cal O}({\cal I}^2).
    \label{s301}
\end{equation}
The terms in the expansion are determined by evaluating the Feynman diagrams in Fig. \ref{1st order diagrams}.   

\subsection{Exact Formula}

The zeroth order diagram, Fig. \ref{1st order diagrams}a leads to the expression in Eq. \ref{s30}.     To first order in $u$, we evaluate the diagram in Fig \ref{1st order diagrams}b in which each of the three density vertices is dressed by a single bubble diagram.   As described in Appendix \ref{appendix a1}, the result can be expressed as
\begin{equation}
s_3^{(1)}(\bq_1,\bq_2) = \sum_{a<b=1}^3 L(\bq_a,\bq_b)
\label{s31}
\end{equation}
with 
\begin{equation}
    L= \frac{2}{N_0} \int \frac{d^2\bk d^2\bp}{(2\pi)^4} \frac{(f_{\bk+\bq_a}+f_{\bk + \bq_b}-1)f_\bk \bar f_{\bk+\bq_{ab}}\bar f_\bp f_{\bp+\bq_{ab}}}{E_{\bk + \bq_{ab}}+E_\bp-E_\bk - E_{\bp+\bq_{ab}}},
    \label{L general}
\end{equation}
where we write $\bq_{ab}=\bq_a+\bq_b$.

In Appendix \ref{appendix a2}, we show that for $E_\bk = |\bk|^2/2m$, the two dimensional integral over $\bp$ can be evaluated analytically.  The result may be written
\begin{equation}\begin{split}
    L(\bq_a,\bq_b) = \int \frac{d^2\bk}{(2\pi)^2}& (f_{\bk+\bq_a}+f_{\bk+\bq_b}-1) f_\bk \bar f_{\bk + \bq_{ab}} \\
  &  \tilde M(\frac{\bk\cdot\bq_{ab}}{k_F|\bq_{ab}|},\frac{|\bq_{ab}|}{2k_F})
  \label{lab main text}
\end{split}\end{equation}
with
\begin{equation}
    \tilde M(u,v) = \frac{1}{\pi v}\left[F(v,u)-F(-v,u+2v)\right].
    \label{m(u,v) main}
\end{equation}
where $F(x,a)$ is evaluated by taking the limit $\lim_{\epsilon\rightarrow 0} {\rm Re}[F(x,a+i\epsilon)]$, with
\begin{equation}
    F(x,a) = a \left(\sqrt{1-\frac{1}{a^2}} \cos^{-1} \frac{1+a x}{a+x} - \cos^{-1} x\right).
    \label{fxa main}
\end{equation}

We have not evaluated the remaining integrals analytically, but it is straightforward to compute them numerically.   Representative cases will be plotted in Section \ref{representative cases}.  In the following section we show that the results simplify in the limit $|\bq_a| \ll k_F$.

\subsection{Long-wavelength limit}

Further analytic progress is possible when $|\bq_1|,|\bq_2| \ll k_F$ (but $\bq_1$ and $\bq_2$ are not necessarily collinear).   In that case, as shown in Appendix \ref{Appendix Analytic Limit}, we obtain
\begin{equation}\begin{split}
    L(\bq_a,\bq_b)& = \frac{|\bq_a||\bq_b|}{(2\pi)^2}\Big(- |\sin(\phi_a+\phi_b)|\\
  &  +\cos\phi_a W(\cos\phi_b)+\cos\phi_b W(\cos\phi_a)\Big),
  \label{L analytic}
\end{split}\end{equation}
where 
\begin{equation}
    \phi_a = \cos^{-1}\left[-\frac{\bq_b\cdot\bq_c}{|\bq_b||\bq_c|}\right]
\end{equation} 
is the internal angle opposite $\bq_a$ in the triangle formed by $\bq_a,\bq_b,\bq_c =-\bq_a-\bq_b$, and
\begin{equation}
    W(u) = \frac{1}{\pi}\left(1+\frac{1-u^2}{2u} \log\frac{1+u}{1-u}\right).
\end{equation}

\begin{figure}
    \centering
    \includegraphics[width=0.75\linewidth]{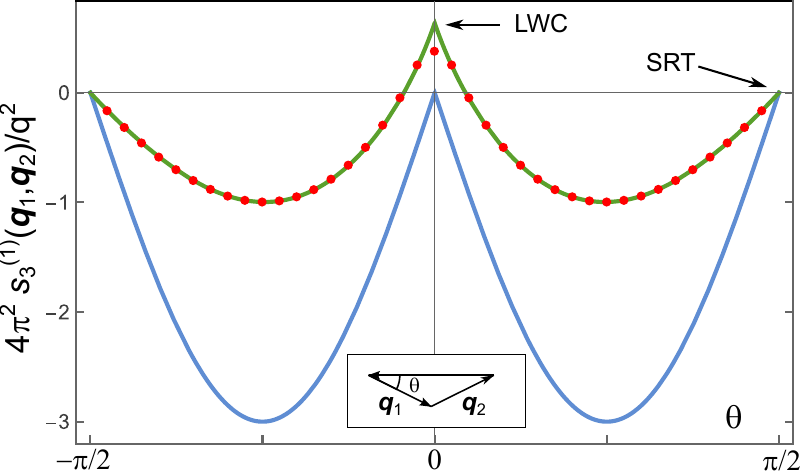}
    \caption{First order perturbative correction to the three-point correlation $s_3^{(1)}(\bq_1,\bq_2)$ for $|\bq_1|=|\bq_2|=.2 k_F$ as a function of the angle $\theta$ between $\bq_1$ and $\bq_2$.  The red dots are the numerical integration of the exact first order correction (\ref{s31},\ref{L general}), and the green curve is the $|\bq_{1,2}|\ll k_F$ asymptotic limit (\ref{L analytic}).  The blue curve is the first order Fermi liquid theory prediction (\ref{s3rho old}), which describes the singularity in the LWC limit.}
    \label{fig2}
\end{figure}

In Fig. \ref{fig2}, we show $s_3^{(1)}(\bq_1,\bq_2)$ for the case in which $\bq_1$, $\bq_2$ and $\bq_3$ form an isosceles triangle with 
$|\bq_1|=|\bq_2| = q =.2 k_F$  with an angle $\theta$ between $\bq_1$ (or $\bq_2$) and $-\bq_3$.   It can be seen that like $s_3^{(0)}(\bq_1,\bq_2)$, $s_3^{(1)}(\bq_1,\bq_2)$ exhibits a cusp-like singularity at $\theta = 0$ and $\theta=\pi/2$.   The limit 
 $\theta \rightarrow 0$ corresponds to the LWC limit in which the Fermi liquid theory prediction in Eq. \ref{s3rho old} is expected to be valid.  In the limit $\theta\rightarrow \pm \pi/2$, $\bq_3 \rightarrow 0$, and this limit is equivalent to that of a skinny right triangle (SRT).

 The blue curve in Fig \ref{fig2} shows the Fermi liquid theory prediction $s_3^{(1)} = -3 |\bq_1\times\bq_2|/(2\pi)^2 = -3 q^2 |\sin 2\theta|/(2\pi)^2$, obtained from \ref{s3rho old}.  The red dots are the result of numerically integrating Eq. \ref{L general}, while the green curve is the $|\bq_{1,2}|\ll k_F$ limit in Eq. \ref{L analytic}.   It can be seen that Eq. \ref{L analytic} is quite accurate even for $q/k_F = .2$ outside a small region near $\theta=0$, where deviations from (\ref{L general}) can be seen.  These deviations occur in the regime where the topological condition (\ref{topological condition}) is violated.   For smaller $q/k_F$, that region of disagreement gets smaller.   It can also be seen that in the LWC limit the $|\theta|$ singularity at $\theta=0$ is correctly predicted by the Fermi liquid theory result, though the cusp is shifted upward by a non-singular constant offset that is not accounted for in the Fermi liquid theory.   Away from the LWC limit, the Fermi liquid theory prediction does not quantitatively reproduce the exact first order perturbation correction, even for $|\bq_1|,|\bq_2| \ll k_F$.

\subsubsection{Long-wavelength collinear limit}

The result in Eq. \ref{L analytic} can be simplified further in the LWC limit, in which $|\bq_a| \ll k_F$, but $\bq_a$ are all nearly parallel (or anti-parallel).   For example, we can choose $\bq_1 = q_1 \hat x + q_\perp \hat y$, $\bq_2 = q_2 \hat x - q_\perp \hat y$, $\bq_3 = -(q_1+q_2)\hat x$, with $0< |q_\perp| \ll q_1, q_2 \ll k_F$.
In that case, we have $|\phi_1|,|\phi_2|,|\pi - \phi_3| \ll 1$, so that (\ref{s31},\ref{L analytic}) reduces to
\begin{equation}
    s_3^{(1)}(\bq_1,\bq_2) = \frac{1}{(2\pi)^2}\left[ -3 (q_1+q_2)|q_\perp| + \frac{2}{\pi} q_1 q_2\right].
    \label{s3 lwc}
\end{equation}
The $|q_\perp|$ singularity at $q_\perp=0$ agrees with the $|\bq_1\times\bq_2|$ singularity predicted by the Fermi liquid theory in Eq. \ref{s3rho old}.   However, there is an additional non-singular offset $q_1 q_2/(2\pi^3)$ that is not accounted for by the Fermi liquid theory.

\subsubsection{Skinny right triangle limit}

An alternative configuration of $\bq_1$ and $\bq_2$, which is useful for interpreting experiments, is to consider $\bq_1 \perp \bq_2$ (or $\bq_1\perp \bq_3$).
For $\bq_1 = q_1 \hat x$ and $\bq_2 = q_\perp \hat y$, the topological condition (\ref{topological condition}) is satisfied for all $q_1$ and $q_\perp$, and allows access to the 
$|q_\perp|$ singularity of $|\bq_1 \times \bq_2|$ even when $|\bq_1|$ is not small.   However,  for $q_\perp\rightarrow 0$ with finite $q_1$ (which defines a skinny right triangle (SRT)), 
the LWC limit does not apply.   Rather, the SRT limit is equivalent to the $\theta=\pi/2$ limit of the isosceles triangle in Fig. \ref{fig2}, in which for
$|q_\perp| \sim q_3 \ll q_1 \ll k_F$, we have $\phi_1 = \phi_2 = \pi/2-\phi/2$, $\phi_3 = \phi\ll 1$.  Then (\ref{s31},\ref{L analytic}) becomes
\begin{equation}
    s_3^{(1)}(\bq_1,\bq_2) =-3\left(1-\frac{2}{\pi}\right) \frac{q_1 |q_\perp|}{(2\pi)^2}.
    \label{s3 srt}
\end{equation}
In this limit there is still a $|\bq_1\times \bq_2|$ type singularity in the first order correction, but its coefficient is reduced by a factor of $1-2/\pi$ from the LWC limit Fermi liquid theory prediction.  The offset is equal to zero.

\subsection{Representative Cases}
\label{representative cases}

In Fig. \ref{Representative Cases} we plot the first order in ${\cal I}$ approximation to $s_3^\rho(\bq_1,\bq_2)$ (\ref{s301}) for a moderately weak attractive interaction, ${\cal I} = -0.2$ for representative configurations of $\bq_1$ and $\bq_2$.   In each of the panels, the gold curve shows the non-interacting (${\cal I}=0$) prediction from Eqs. \ref{top formula} and \ref{i=0 formula}.   The red dotted line shows the result of the exact first order theory which follows from ({\ref{L general}) by numerically integrating (\ref{lab main text}).  The green curves show the analytic formula (\ref{L analytic}), derived for $|\bq_1|,|\bq_2|\ll k_F$.   The blue line shows the Fermi liquid theory prediction (\ref{s3rho old}), which is expected to match the singularity in the LWC limit.

\begin{figure}
    \centering
    \includegraphics[width=1\linewidth]{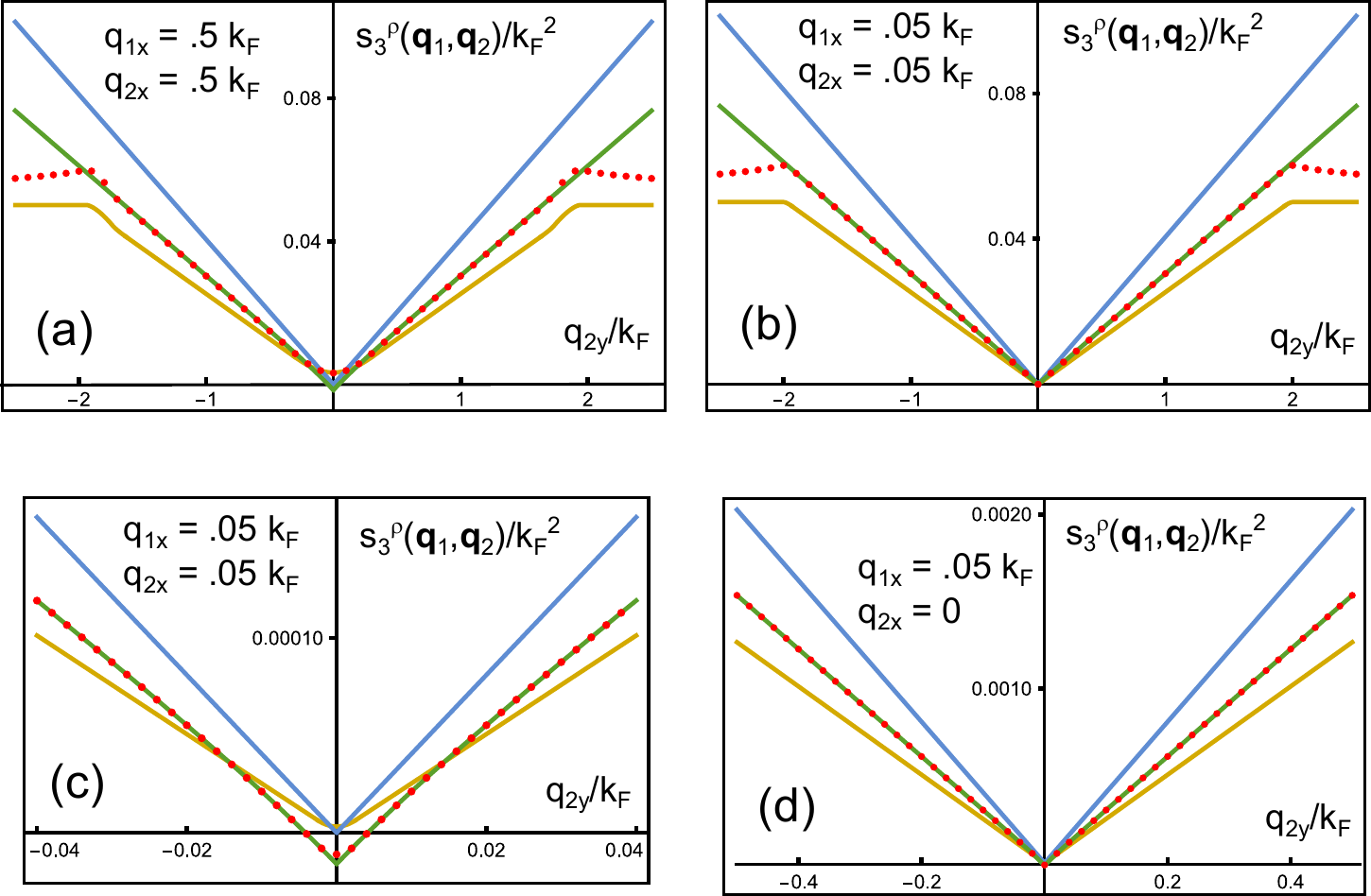}
    \caption{Three-point total density correlation function $s_3^\rho(q_{1x}\hat x,q_{2x}\hat x + q_{2y}\hat y)$ as a function of $q_{2y}$ for different configurations of fixed $q_{1x}$ and $q_{2x}$.  The gold lines show the exact non-interacting prediction (\ref{top formula},\ref{i=0 formula}), while the other data show approximations for a moderately weak attractive interaction ${\cal I}=-0.2$.  The blue line shows the Fermi liquid theory prediction (\ref{s3rho old}) extrapolated beyond the LWC limit.   The red dots are the numerical integration of the exact first order theory (\ref{L general},{\ref{lab main text}}), while the green lines are the analytic formula (\ref{L analytic}) derived for $|\bq_1|,|\bq_2|\ll k_F$.    } 
    \label{Representative Cases}
\end{figure}

Fig. \ref{Representative Cases}a shows $s_3^\rho(\bq_1,\bq_2)$ for $\bq_1 = 0.5 k_F \hat x$, $\bq_2 = 0.5 k_F \hat x + q_{2y}\hat y$ as a function of $q_{2y}$.  It can be seen that the $|\bq_1\times \bq_2|$ singularity is rounded for small $|q_{2y}|$, due to the violation of the topological condition (\ref{topological condition}), which in the present case is $q_{2y} > 0.307 k_F$.   For larger $q_{2y}$ the linear behavior resembles the topological formula with a slope that is modified by interactions.   However, the slope is different from the coefficient of the singularity predicted in the LWC limit.

Fig. \ref{Representative Cases}b shows $s_3^\rho(\bq_1,\bq_2)$ for $\bq_1 = 0.05 k_F \hat x$, $\bq_2 = 0.05 k_F \hat x + q_{2y}\hat y$ as a function of $q_{2y}$.   In this case, the $|\bq_1\times\bq_2|$ singularity is sharp because the topological regime (\ref{topological condition}) remains valid for much smaller $q_{2\perp} > 0.0025 k_F$.   Fig. \ref{Representative Cases}c shows the same configuration over a smaller range of $q_{2y}$, that shows the rounding.   When $q_{2y} \gtrsim .0025 k_F$, the conditions for the LWC limit are satisfied.   It can be seen that the slope of the singularity agrees with the LWC theory (the blue line), but there is an offset, as predicted by (\ref{s3 lwc}).

Fig. \ref{Representative Cases}d shows $s_3^\rho(\bq_1,\bq_2)$ for $\bq_1 = 0.05 k_F \hat x$, $\bq_2 =  q_{2y}\hat y$ as a function of $q_{2y}$.  In this case $\bq_{1,2,3}$ form a right triangle, and the topological condition (\ref{topological condition}) is satisfied for all $q_{2y}$.   In this case, the linear slope is predicted by (\ref{s3 srt}), and differs from the slope in the LWC limit, given by (\ref{s3rho old}).

It is striking that the asymptotic formula (\ref{L analytic}) gives results very close to the numerical integration of (\ref{lab main text}) when the topological condition (\ref{topological condition}) is satisfied, even when $|\bq_{1,2}|$ are not much less than $k_F$.  However, there are slight numerical deviations between the two results, so it is unlikely that (\ref{L analytic}) is exact away from the small $\bq_{1,2}$ limit.

\section{Third Order Correction for Same-Spin Correlations}

\label{Section 4}

In this section, we compute the leading order correction to the same-spin three-point density correlation function in the LWC limit using (\ref{s3up}) and determining the necessary corrections to the Landau parameters.   To this end, it is useful to express (\ref{s3up}) in terms of the Landau parameters for same-spin (opposite-spin) interactions:
\begin{align}
    F_0^s &= F_0^{\uparrow\uparrow} + F_0^{\uparrow\downarrow}\\
    F_0^a &= F_0^{\uparrow\uparrow} - F_0^{\uparrow\downarrow}.
\end{align}
Then,
\begin{equation}
    s_3^\uparrow(\bq_1,\bq_2) = \frac{|\bq_1\times\bq_2|}{(2\pi)^2}\frac{(1+F_0^{\uparrow\uparrow})^3- (F_0^{\uparrow\downarrow})^3}{\left((1+F_0^{\uparrow\uparrow})^2 - (F_0^{\uparrow\downarrow})^2\right)^3}.
\end{equation}
We now consider a weak contact interaction between up and down spins, and expand in powers of the interaction parameter ${\cal I}$.    Noting that for a contact interaction, the same-spin Landau parameter $F_0^{\uparrow\uparrow}$ vanishes to linear order in ${\cal I}$ we have
\begin{align}
    F_0^{\uparrow\downarrow} &= a_1 {\cal I} + a_2 {\cal I}^2 + a_3 {\cal I}^3 + {\cal O}({\cal I}^4),\\
   F_0^{\uparrow\uparrow} &= b_2 {\cal I}^2 + b_3 {\cal I}^3 + {\cal O}({\cal I}^4). 
\end{align}
Then
\begin{equation}\begin{split}
      s_3^\uparrow(\bq_1,\bq_2) = &\frac{|\bq_1\times\bq_2|}{(2\pi)^2} \Big( 1 + 3 ( a_1^2-b_2){\cal I}^2 \\
   &   + (-a_1^3+6a_1 a_2-3 b_3){\cal I}^3 + {\cal O}({\cal I}^4)\Big).
\end{split}\end{equation}
$a_1$, $a_2$ and $b_2$ were computed earlier\cite{Randeria1992}, and are equivalent to (\ref{f0s,f0a}):   
\begin{equation}
    a_1 = 1 ; \quad\quad
    a_2 = 1- \log 2;\quad\quad
    b_2 = 1.
    \label{a1 a2 b2 eq}
\end{equation}
For completeness, those results are reproduced below.   Eq. \ref{a1 a2 b2 eq} leads to the non-trivial prediction found in Ref. \cite{TamKane2026} that the ${\cal O}({\cal I}^2)$ contribution to $s_3^\uparrow$ vanishes.   To determine the leading ${\cal O}({\cal I}^3)$ contribution, note that it only depends on $b_3$, and not on $a_3$.   Thus, the only calculation required is to determine the third order contribution to the same-spin Landau parameter $F_0^{\uparrow\uparrow}$.
The rest of this section is devoted to that cumbersome calculation.   The result, however, is quite simple.  We find
\begin{equation}
    b_3 = 0.
    \label{b3 eq}
\end{equation}
It follows that to ${\cal O}({\cal I}^3)$ in the LWC limit, 
\begin{equation}
    s^\uparrow_3(\bq_1,\bq_2) = \frac{|\bq_1\times\bq_2|}{(2\pi)^2} \left[1 + (5 - 6 \log 2) {\cal I}^3  + {\cal O}({\cal I}^4)\right].
\end{equation}
This is the central result of this section.

\subsection{Perturbative expansion of Landau Parameters}

We now perform the perturbative expansion of the Landau parameters to verify Eq. \ref{a1 a2 b2 eq} and to establish Eq. \ref{b3 eq}.   As noted in the previous section, we need to compute $F_0^{\uparrow\uparrow}$ to ${\cal O}({\cal I}^3)$, but $F_0^{\uparrow\downarrow}$ only needs to be computed to ${\cal O}({\cal I}^2)$.

There are two methods for computing the Landau parameters.   Landau's original method was to consider a formal expansion in powers of the interaction of the ground state energy $E_0(\{n_{\bk,\sigma}\})$ as a function of the fermion occupation numbers near $n_{\bk,\sigma}$ the Fermi energy\cite{agd,landau1957theory}.   The Landau interaction is then obtained by differentiating,
\begin{equation}
    f_{\bk\bk'}^{\sigma\sigma'} = \frac{\partial^2 E_0}{\partial n_{\bk\sigma}\partial n_{\bk'\sigma'}}.
\end{equation}
An alternative, and equivalent approach is to consider a diagrammatic expansion for the single-particle propagator\cite{agd,landau1959theory}.   The Landau parameter is then expressed in terms of the exact two-particle irreducible vertex function evaluated on the energy shell $\omega = E_{\bk} = E_F$ for external energy and momentum in the particle-hole channel in the limit $\Omega\rightarrow 0$, $|\bq|\rightarrow 0$, $|\bq|/\Omega \rightarrow 0$:
\begin{equation}
     f_{\bk\bk'}^{\sigma\sigma'} = Z^2 \Gamma^{\sigma\sigma'}_{\bk\bk'}.
\end{equation}
The quasiparticle residue $Z$ is expressed in terms of the exact self-energy $\Sigma(\bk,\omega)$ as
\begin{equation}
    Z = \frac{1}{1-\partial\Sigma/\partial \omega}.
\end{equation}

\begin{figure}
    \centering
    \includegraphics[width=.9\linewidth]{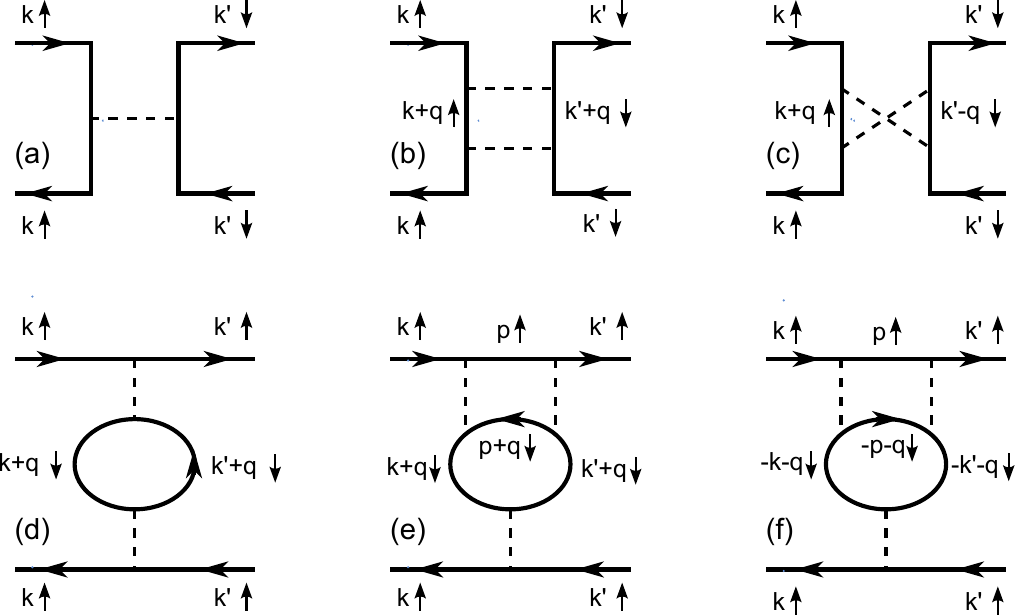}
    \caption{Feynman diagrams for the vertex function $\Gamma^{\uparrow\downarrow}_{\bk\bk'}$ (a,b,c) and $\Gamma^{\uparrow\uparrow}_{\bk\bk'}$ (d,e,f).  The diagrams are to be evaluated with the four external legs removed.  (e) and (f) are counted twice to account for diagrams related by reflection.}
    \label{Gamma diagrams}
\end{figure}

We have performed the calculation using both methods and checked that they agree and lead to Eqs. \ref{fupdown} and \ref{fupup}.   Here we will report the diagrammatic calculation and compute $\Gamma_{\bk\bk'}^{\uparrow\downarrow}$  ($\Gamma_{\bk\bk'}^{\uparrow\uparrow}$) to second (third) order in an expansion in powers of the coefficient $u$ of a $\delta$-function contact interaction between $\uparrow$ and $\downarrow$ spins.   As mentioned earlier, and noted by Engelbrecht et al.\cite{Randeria1992} using the method of Abrikosov and Kalatnikov\cite{AbrikosovKhalatnikov1957}, this expansion is divergent in the ultraviolet, but that divergence can be repaired by re-expressing it in terms of the renormalized interaction ${\cal I}$, which characterized the physical scattering length for particles at the Fermi energy.   We will regularize the divergences with an ultraviolet cutoff, which allows us to determine the non-divergent coefficients at each order in perturbation theory in ${\cal I}$.

Note that for the contact interaction, the only contribution to the self energy $\Sigma$ to linear order in $u$ is the constant Hartree term, which just shifts the zero in the energy.   Thus, the quasiparticle residue will be $Z = 1 + {\cal O}(u^2)$.   This means that to the order that is required we may replace $Z$ by 1.    

The necessary vertex functions are described by the Feynman diagrams in Fig. \ref{Gamma diagrams}.   Some details of their evaluation are included in Appendix \ref{appendix b1}.   Fig. \ref{Gamma diagrams}a,b,c lead to
\begin{equation}\begin{split}
    f^{\uparrow\downarrow}_{\bk\bk'} = u - u^2 &\int \frac{d^2\bq}{(2\pi)^2}\Bigg[ \frac{f_{\bk'+\bq}-f_{\bk+\bq}}{E_{\bk'+\bq}-E_{\bk+\bq}} \\
    &+ \frac{f_{\bk'-\bq}+f_{\bk+\bq}-1}{2E_F - E_{\bk'-\bq}-E_{\bk+\bq}}\Bigg] + {\cal O}(u^3)
    \end{split}
    \label{fupdown}
\end{equation}
and Fig. \ref{Gamma diagrams}d,e,f lead to
\begin{equation}\begin{split}
    f^{\uparrow\uparrow}_{\bk\bk'} =  &- u^2\int \frac{d^2\bq}{(2\pi)^2} \frac{f_{\bk'+\bq}-f_{\bk+\bq}}{E_{\bk'+\bq}-E_{\bk+\bq}} 
    + u^3 \int \frac{d^2\bp d^2\bq}{(2\pi)^4} \\
   &  \Bigg[ \frac{ 2(f_\bp \bar f_{\bp+\bq} f_{\bk+\bq}+\bar f_\bp  f_{\bp+\bq} \bar f_{\bk + \bq})}{(E_{\bk'+\bq}-E_{\bk+\bq})(
    E_F + E_{\bp+\bq}-E_\bp-E_{\bk+\bq})} \\
   + &  \frac{ 2(f_\bp f_{\bp+\bq}\bar f_{\bk+\bq}+\bar f_\bp \bar f_{\bp+\bq} f_{\bk+\bq})}{(E_{\bk'+\bq}-E_{\bk+\bq})(
    E_F + E_{\bk+\bq}-E_\bp-E_{\bp+\bq})} \\
    + &\bk \rightleftharpoons \bk' \Bigg] + {\cal O}(u^4).
\end{split}
\label{fupup}
\end{equation}
The  dimensionless Landau parameters are then given by
\begin{equation}
    F_0^{\sigma\sigma'} = \left\langle N_0 f_{\bk\bk'}^{\sigma\sigma'} \right\rangle_{\varphi,\varphi'} \equiv N_0\int \frac{d\varphi d\varphi'}{(2\pi)^2}  f^{\sigma\sigma'}_{\bk \bk'},
    \label{F0 avarage}
\end{equation}
where $\bk = k_F (\cos\varphi,\sin\varphi)$ and $\bk'=k_F(\cos\varphi',\sin\varphi')$.   Remarkably, as outlined in Appendix \ref{Appendix B},
the integrals can be evaluated analytically, leading to 
\begin{equation}\begin{split}
   & F_0^{\uparrow\downarrow} = N_0 u + (N_0 u)^2 (I_1 + I_2) + {\cal O}(u^3),\\
  &  F_0^{\uparrow\uparrow} = (N_0 u)^2 I_1 + (N_0 u)^3 (I_3 + I_4) + {\cal O}(u^4),
  \label{F0I}
\end{split}\end{equation}
where $I_1$ and $I_2$ follow from the evaluation the first and second lines of (\ref{fupdown}), which represent Figs. \ref{Gamma diagrams}(b,d) and Fig. \ref{Gamma diagrams}c, while
$I_3$ and $I_4$ follow from the evaluation of the second and third lines of (\ref{fupup}), which represent Fig. \ref{Gamma diagrams}e and Fig. \ref{Gamma diagrams}f.    The resulting integrals for $I_{1,2,3,4}$ are listed in Appendix \ref{appendix b2} and are evaluated in Appendix \ref{appendix b3}.   We find
\begin{equation}\begin{split}
    I_1 &= 1, \\
    I_2 &= - \log \frac{k_\Lambda}{k_F}- \log 2 ,\\
    I_3 &= 2(1-\log 2),\\
    I_4 &= -2 \log \frac{k_\Lambda}{k_F} - 2 (1- \log 2).
    \label{i results}
\end{split}\end{equation}
The divergent dependence on the ultraviolet cutoff $k_\Lambda$ is repaired by using (\ref{u to i}) to express the expansion in terms of
the renormalized interaction parameter ${\cal I}$, which leads to
\begin{align}
  &  F_0^{\uparrow\downarrow} = {\cal I} + {\cal I}^2 (1- \log 2) + {\cal O}({\cal I}^3), \\
  &  F_0^{\uparrow\uparrow} = {\cal I}^2 + {\cal O}({\cal I}^4).
\end{align}
The expression for $F_0^{\uparrow\downarrow}$ reproduces the known\cite{Randeria1992} second order expansion of the Landau parameters (\ref{f0s,f0a}).  The non-trivial new result is the vanishing of the third order correction to $F_0^{\uparrow\uparrow}$.

\section{Conclusion}

In this paper we have extended the calculation of the three-point density correlations of two dimensional fermions interacting via a contact interaction in two ways.   We have computed the correlations exactly to first order in perturbation theory in the dimensionless interaction parameter ${\cal I}$.    The ${\cal O}({\cal I})$ term vanishes for the same-spin correlations $s_3^\uparrow(\bq_1,\bq_2)$, but for the total density correlations $s_3^\rho(\bq_1,\bq_2)$ it is a non-trivial function of $\bq_1$ and $\bq_2$.   In the long-wavelength collinear (LWC) limit, in which $\bq_1$ and $\bq_2$ are nearly parallel, the results agree with our earlier theory, which expressed the coefficient of the $|\bq_1\times\bq_2|$ singularity of the three-point correlations in that limit in terms of the Landau Fermi liquid parameters $F_0^s$ and $F_0^a$.    The exact first order calculation applies to general values of $\bq_1$ and $\bq_2$, and is valid beyond the LWC limit.

Our second result concerned the leading order contribution to the same-spin correlations $s_3^\uparrow(\bq_1,\bq_2)$ in the LWC limit.   Our previous theory predicted that the correction vanished to order ${\cal O}({\cal I}^2)$.   Here, we have shown that the ${\cal O}({\cal I}^3)$ correction does not vanish, and we computed its value.   This involved a non-trivial diagrammatic computation of the third order correction to the same-spin Landau parameter $F_0^{\uparrow\uparrow}$, which turned out to be equal to zero.   This allowed us to apply our previous result expressing $s_3^\uparrow$ in the LWC limit in terms of the Landau parameters to determine the leading ${\cal O}({\cal I}^3)$ contribution.

Recent experiments have measured the same-spin correlations $s_3^\uparrow(\bq_1,\bq_2)$ for a Fermi gas of $^6$Li atoms with an attractive contact interaction $-.5 < {\cal I} <0$ that are controlled by tuning a Feshbach resonance\cite{daix2025probing}.   The observed correlations were well described by the non-interacting (${\cal I}=0$) theory, which appears to be consistent with the vanishing of $s_3^\uparrow$ to ${\cal O}({\cal I}^2)$.   The present work suggests that the effects of interactions should be observable in the same-spin correlations for stronger interactions.   While stronger attractive interactions ${\cal I} \lesssim -1$ are complicated by the BCS pairing instability, the Fermi liquid theory developed here also applies to repulsive interactions ${\cal I}>0$, where the Fermi liquid theory is expected to apply for stronger interactions.    In addition, our results allow for a meaningful comparison between experiment and theory for the three-point correlations of the total density $s_3^\rho(\bq_1,\bq_2)$ even in the regime $-.5<{\cal I}<0$.   This motivates further experimental studies in that direction.

\acknowledgments
We thank Pok Man Tam for discussions and collaboration that formed the foundation for this work.   We also thank Tarik Yefsah, Bruno Peaudecerf and Cyprien Daix for discussions of the experiments that motivated this work.

\bibliography{refs}

\appendix

\section{First order for $s_3^\rho$}
\label{Appendix A}

In this appendix we provide some details for the evaluation of the first order correction to the total density three-point correlation function.  

\subsection{Evaluation of Feynman Diagrams}
\label{appendix a1}

It is simplest to evaluating the diagrams in Fig. \ref{1st order diagrams}b in the time domain using a zero temperature Euclidean time formulation.   Evaluating the diagram 
for the equal-time correlator, along with a similar one with $a$ and $b$ exchanged gives
\begin{equation}\begin{split}
    u &\int \frac{d^2\bk d^2\bp}{(2\pi)^4} \int_{-\infty}^\infty  d\tau \left[G_\bp(\tau)G_{\bp+\bq_{ab}}(-\tau)\right] \\
   & \left( G_\bk(-\tau)[ G_{\bk+\bq_a}(0^+) + G_{\bk + \bq_b}(0^-)]G_{\bk+\bq_{ab}}(\tau) \right),
   \label{ggggg}
\end{split}\end{equation}
where use the shorthand $\bq_{ab}=\bq_a+\bq_b$ and we have specified a small time difference between $\rho(\bq_a)$ and $\rho(\bq_b)$ to fix the operator order.   The single fermion propagator is
\begin{equation}
    G_\bk(\tau) = \int \frac{d\omega}{2\pi} \frac{e^{-i \omega\tau}}{i\omega - E_{\bk}} = (f_\bk - \theta(\tau)) e^{-E_\bk \tau},
    \label{gktau}
\end{equation}
where $E_\bk = |\bk|^2/(2m)$, $f_\bk = \theta(E_F-E_\bk)$ is the zero temperature Fermi occupation, and $\theta$ is a step function.   Eq. \ref{L general} is then recovered by substituting $u \rightarrow {\cal I}/N_0$ and plugging (\ref{gktau}) into (\ref{ggggg}), noting that the integrals for $\tau>0$ and $\tau<0$ are identical after a substitution $\bk \rightarrow \bq_{ab}-\bk$ and $\bp \rightarrow \bq_{ab}-\bp$ and using $E_{-\bk}=E_\bk$, which accounts for the factor of $2$.

\subsection{Evaluation of integrals}
\label{appendix a2}
\begin{figure}
    \centering
    \includegraphics[width=0.9\linewidth]{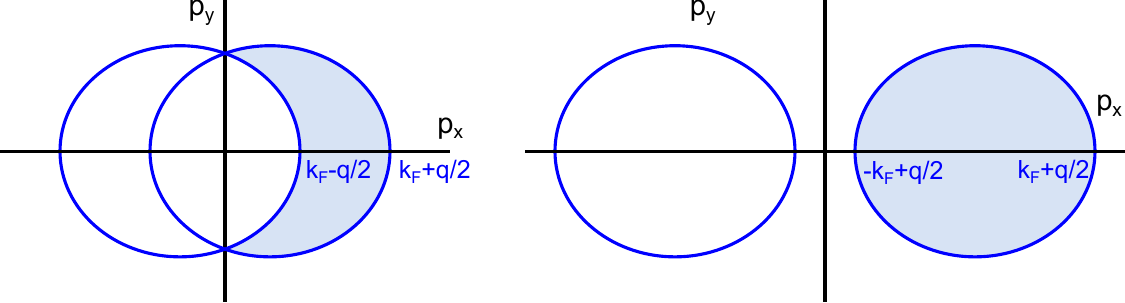}
    \caption{Region of integration for $\bar f_{\bp+\bq/2} f_{\bp-\bq/2}$ for (a) $|\bq|<2k_F$ and (b) $|\bq|>2k_F$.}
    \label{regionfig}
\end{figure}

To simplify the integral in Eq. \ref{L general}, we write
\begin{equation}
    L(\bq_a,\bq_b) = \int \frac{d^2\bk}{(2\pi)^2} (f_{\bk+\bq_a}+f_{\bk+\bq_b}-1) f_\bk \bar f_{\bk + \bq_{ab}} M(\bk ,\bq_{ab}),
    \label{lab appendix}
\end{equation}
where
\begin{equation}
    M(\bk,\bq) = \frac{2}{N_0} \int \frac{d^2\bp }{(2\pi)^2} \frac{\bar f_{\bp} f_{\bp+\bq}}{E_{\bk+\bq_{ab}}+E_\bp-E_\bk-E_{\bp+\bq_{ab}}}.
\end{equation}
Using $E_\bk = |\bk|^2/(2m)$ and $N_0 = m/(2\pi)$ and letting $\bp \rightarrow -\bp-\bq/2$, this becomes
\begin{equation}
    M(\bk,\bq) = \frac{1}{\pi}\int d^2\bp \frac{\bar f_{\bp+\bq/2} f_{\bp-\bq/2}}{\bq \cdot(\bk + \bp+\bq/2)}.
\end{equation}
To evaluate this integral we can, without loss of generality choose $\bq = q \hat \bq$ with $\hat \bq = \hat x$.   Then $M(\bk,\bq)$ depends only on $q$ and $k_x = \bk\cdot\hat \bq$.   The integral over the shaded region in Fig. \ref{regionfig} is independent of $p_y$, and is given by
\begin{equation}\begin{split}
    M = \frac{\theta(2k_F-q)}{\pi q}\Bigg[ &\int_{0}^{k_F+q/2} dp_x \frac{2\sqrt{k_F^2-(p_x-q/2)^2}}{p_x+k_x+q/2}  \\
    - &\int_{0}^{k_F-q/2} dp_x \frac{2\sqrt{k_F^2-(p_x+q/2)^2}}{p_x+k_x+q/2}\Bigg] \\
    +\frac{\theta(q-2k_F)}{\pi q}& \int_{q/2-k_F}^{q/2+k_F} dp_x \frac{2 \sqrt{k_F^2-(p_x-q/2)^2}}{p_x+k_x+q/2}.
\end{split}\end{equation}
Introducing dimensionless variables $u =\bk\cdot\hat\bq/k_F$ and $v = |\bq|/(2k_F)$ and shifting the integration variables, this may be written as
\begin{equation}
    M(\bk,\bq) = \tilde M\left(\frac{\bk\cdot\hat\bq}{k_F},\frac{|\bq|}{2 k_F}\right),
    \label{mtilde appendix}
\end{equation} 
with
\begin{equation}\begin{split}
  \tilde  M(u,v) = \frac{\theta(1-v)}{\pi v} \Bigg[&
    \int_{-v}^1 dx \frac{\sqrt{1-x^2}}{x+u+2v} \\
    -& \int_{v}^1 dx \frac{\sqrt{1-x^2}}{x+u}\Bigg] \\
    +\frac{\theta(v-1)}{\pi v}& \int_{-1}^1 dx \frac{\sqrt{1-x^2}}{x + u+2v}.
    \label{muvintegrals}
\end{split}\end{equation}
These integrals can be evaluated with the aid of the following indefinite integral,
\begin{equation}
 \int dx \frac{\sqrt{1-x^2}}{x+a}   = \sqrt{1-x^2} + F(x,a)
 \label{fintegral}
\end{equation}
with
\begin{equation}
    F(x,a) = a \left(\sqrt{1-\frac{1}{a^2}} \cos^{-1} \frac{1+a x}{a+x} - \cos^{-1} x\right).
    \label{fxa}
\end{equation}
The branch cut in $F(x,a)$ must be treated carefully.   For real $x$ and $a$, the principal part of the integral in (\ref{fintegral}) is
evaluated by giving $a$ a small imaginary part.   Thus $F(x,a)$ should be interpreted as $\lim_{\epsilon\rightarrow 0} {\rm Re}[F(x,a+i\epsilon)]$.
Though in (\ref{muvintegrals}) only $|x|\le 1$ is required, the formulas below are simplified by considering the analytic continuation to $|x|>1$.   
For real $x$, $a$, (\ref{fxa}) is given explicitly by,
\begin{equation}
    \begin{split}
       & \lim_{\epsilon\rightarrow 0} {\rm Re}[F(x,a+i\epsilon)] = \\ 
       &\left\{\begin{array}{cc}
         a \left( \sqrt{1-\frac{1}{a^2}}\cos^{-1}\frac{1+ a x}{a+x}-\cos^{-1}x\right) & |x|<1, |a|>1 \\
         - \sqrt{1-a^2} \cosh^{-1} |\frac{1+a x}{a+x}|-a \cos^{-1} x &  |x|<1,|a|<1 \\
         a \pi\left(\theta(-\frac{x+a}{1+x a})\sqrt{1-\frac{1}{a^2}}-\theta(-x)\right) & |x|>1,|a|>1 \\
       - a \pi \theta(-x) & |x|>1,|a|<1.
\end{array}\right. 
\label{F(x,a)}
    \end{split}
\end{equation}
Then, noting that the first term in (\ref{fintegral}) cancels, we have
\begin{equation}
    \begin{split}
        \tilde M(u,v) = \frac{\theta(1-v)}{\pi v}\Big(&F(1,u+2v)-F(-v,u+2v) \\
        & -F(1,u)+F(v,u)\Big)\\
        + \frac{\theta(v-1)}{\pi v} \big( & F(1,u+2v)-F(-1,u+2v)\big).
    \end{split}
\end{equation}
We next use the fact that $F(1,a)=0$.   In addition, we have checked that the term for $v>1$ has the same functional form as the analytic continuation of the term for $v<1$, so the result can be expressed in the simplified form,
\begin{equation}
    \tilde M(u,v) = \lim_{\epsilon\rightarrow 0}{\rm Re}\left[\frac{F(v,u+i\epsilon)-F(-v,u+2v+i\epsilon)}{\pi v}\right].
    \label{m(u,v)}
\end{equation}
Combining (\ref{lab appendix}), (\ref{mtilde appendix}), (\ref{fxa}) and (\ref{m(u,v)}) then leads to (\ref{lab main text}), (\ref{m(u,v) main}) and (\ref{fxa main}) in the main text.

\subsection{Analytic limit $|\bq_a|\ll k_F$}
\label{Appendix Analytic Limit}
\begin{figure}
    \centering
    \includegraphics[width=0.5\linewidth]{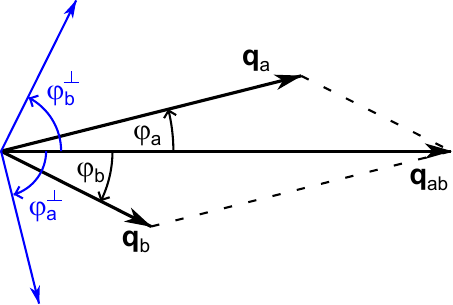}
    \caption{Definition of angles characterizing $\bq_a$, $\bq_b$ and $\bq_{ab}=\bq_a+\bq_b$.   For this configuration $\varphi_a,\varphi_b^\perp>0$ and $\varphi_b,\varphi_a^\perp<0$.}
    \label{trianglefig}
\end{figure}
We now consider the limit $\bq \rightarrow 0$.   In this case we can consider $\tilde M_0(u) = \lim_{v\rightarrow 0} \tilde M(u,v)$, given by
\begin{equation}
\tilde M_0(u) = 1 - \frac{2}{\pi} \frac{u}{\sqrt{1-u^2}}\cosh^{-1}\frac{1}{|u|}.
\label{m0(u)}
\end{equation}
It is useful to rewrite the other term in the integrand in (\ref{lab appendix}) in the following form:
\begin{equation}
\begin{split}
    (f_{\bk+\bq_a}+f_{\bk+\bq_b}&-1) f_\bk \bar f_{\bk + \bq_{ab}} = \\
    - \theta(\bv_\bk\cdot \bq_{ab}) 
   \Big\{&\bq_b \cdot\nabla_\bk\left[  s(\bv_\bk \cdot \bq_a) \bq_a \cdot \nabla_\bk f_\bk \right] \\
    +& \bq_a \cdot\nabla_\bk\left[ s(\bv_\bk \cdot \bq_b) \bq_b \cdot \nabla_\bk f_\bk \right] \Big\},
\end{split}
\label{derivative formula}
\end{equation}
where we write $s(x) = \theta(x)-1/2$.
Eq. \ref{derivative formula} can be derived by repeatedly applying the identities $f_{\bk + \bq}\bar f_\bk = \theta(E_{\bk+\bq}-E_\bk) (f_\bk - f_{\bk + \bq})
=\theta(E_{\bk+\bq}-E_\bk) (\bar f_{\bk + \bq}-\bar f_\bk)$ to the left hand side and taking the $\bq\rightarrow 0$ limit.

Plugging (\ref{m0(u)}) and (\ref{derivative formula}) into (\ref{lab appendix}) and integrating by parts, we obtain
\begin{equation}\begin{split}
   & L = \int \frac{d^2\bk}{(2\pi)^2} \Big[\\
    &(\bq_a\cdot\nabla_\bk f_\bk)s(\bv_\bk \cdot \bq_a) \bq_b\cdot\nabla_\bk\left(\theta(\bv_\bk\cdot \bq_{ab}) \tilde M^0(u_\bk)\right) \\
    +& (\bq_b\cdot\nabla_\bk f_\bk)s(\bv_\bk \cdot \bq_b) \bq_a\cdot\nabla_\bk\left(\theta(\bv_\bk\cdot \bq_{ab}) \tilde M^0(u_\bk)\right)\Big],
\end{split}\end{equation}
where we write $u_\bk = \bk \cdot \hat\bq_{ab}/k_F$.

It can be seen that the presence of $\nabla_\bk f_\bk$ restricts the integration to the Fermi surface $|\bk|=k_F$.    We can separate $L = L_0 + L_1$ into two terms that depend on which factor in the large parentheses the second $\nabla_\bk$ acts on.  For $L_0$, the derivative acts on $\theta(\bv_k\cdot\bq_{ab})$, which restricts the integral to the critical point on the Fermi surface for which $\bv_\bk \perp \bq_{ab}$.   At that point, $\tilde M_0(u_\bk) = \tilde M_0(0) = 1$.   The $L_0$ integral with $\tilde M_0=1$ is identical to (minus) the integral in Eq. \ref{s30} for $s_3^{(0)}$, and is given by
\begin{equation}
    L_0 = - \frac{|\bq_1\times \bq_2|}{(2\pi)^2}.
\end{equation}

The remaining integral is
\begin{equation}
\begin{split}
L_1 = &\int \frac{d^2\bk}{(2\pi)^2} \theta(\bv_\bk \cdot \bq_{ab}) \Big[ \\
&(\bq_a \cdot \nabla_\bk f_\bk) s(\bv_\bk\cdot \bq_a) \bq_b \cdot\nabla \tilde M_0(u_\bk) \\
+& (\bq_b \cdot \nabla_\bk f_\bk) s(\bv_\bk\cdot \bq_b) \bq_a \cdot\nabla \tilde M_0(u_\bk)\Big].
\end{split}
\end{equation}
The integral can be performed using polar coordinates $\bk = k( \hat \bq_{ab}\cos\varphi +  \hat z \times\hat \bq_{ab}\sin\varphi)$.   Then, $(\bq_a\cdot\nabla_\bk f_\bk) s(\bv_\bk\cdot\bq_a) = - |\bq_a| |\cos(\varphi-\varphi_a)|\delta(k - k_F)/2$ and $\bq_b \cdot \nabla_\bk \tilde M_0(u_\bk) = |\bq_b|\cos \varphi_b \tilde M_0'(\cos\varphi)/k_F$, where $\varphi_{a}$ is the angle between $\bq_{a}$ and $\bq_{ab}$ indicated in Fig. \ref{trianglefig}.  We then have
\begin{equation}\begin{split}
    L_1 &= -\frac{|\bq_a||\bq_b|}{8\pi^2}\int_{-\pi/2}^{\pi/2} d\varphi \tilde M_0'(\cos\varphi)\\
    &\big[|\cos(\varphi-\varphi_a)| \cos\varphi_b - |\cos(\varphi-\varphi_b)|\cos\varphi_a\big].
    \label{l1 abs}\end{split}\end{equation}
To evaluate the absolute value, define $\varphi_a^\perp \in [-\pi/2,\pi/2]$ to be the direction perpendicular to $\bq_a$:
$\varphi_a^\perp = \varphi_a - (\pi/2)\sgn\varphi_a$.   Then, $\sgn[\cos(\varphi-\varphi_a)] = \sgn[\varphi_a(\varphi-\varphi_a^\perp)]$ (see Fig. \ref{trianglefig}).  Then,
\begin{equation}\begin{split}
    L_1 &= -\frac{|\bq_a||\bq_b|}{8\pi^2}\int_{-\pi/2}^{\pi/2} d\varphi \tilde M_0'(\cos\varphi)\\
    &\big[\sgn\varphi_a\sgn(\varphi-\varphi_a^\perp)\cos(\varphi-\varphi_a)\cos\varphi_b \\
    &+ \sgn\varphi_b\sgn(\varphi-\varphi_b^\perp)\cos(\varphi-\varphi_b)\cos\varphi_a\big].
\end{split}\end{equation}
\begin{equation}
    \begin{split}
        = -\frac{|\bq_a||\bq_b|}{8\pi^2} \Bigg\{&\int_{-\pi/2}^{\pi/2}d\varphi \tilde M_0'(\cos\varphi)\sin\varphi\Big[ \\
       & \sgn\varphi_a\sgn(\varphi-\varphi_a^\perp)\sin\varphi_a\cos\varphi_b\\
        +&\sgn\varphi_b\sgn(\varphi-\varphi_b^\perp)\cos\varphi_a\sin\varphi_b\Big] \\
       +&\int_{-\pi/2}^{\pi/2}d\varphi \tilde M_0'(\cos\varphi)\cos\varphi\Big[ \\
          & \sgn\varphi_a\sgn(\varphi-\varphi_a^\perp)\cos\varphi_a\cos\varphi_b\\
        +&\sgn\varphi_b\sgn(\varphi-\varphi_b^\perp)\cos\varphi_a\cos\varphi_b\Big] \Bigg\}.
    \end{split}
\end{equation}
The integrals can now be performed by applying the appropriate limits to the indefinite integrals
\begin{equation}
    \int d\varphi \tilde M_0'(\cos\varphi)\sin\varphi = - \tilde M_0(\cos\varphi)
\end{equation}
and
\begin{equation}
\tilde D_0(\cos\varphi) \equiv \int d\varphi\tilde M_0'(\cos\varphi)\cos\varphi. 
\end{equation}
It can be checked that
\begin{equation}
    \tilde D_0(u) = \frac{1}{\pi}\left[\frac{1}{\sqrt{1-u^2}}- \frac{u^2}{1-u^2}\cosh^{-1}\frac{1}{|u|}\right]
    \label{d0(u)}
\end{equation}
satisfies $\tilde D_0'(u) = \tilde M_0'(u) u/\sqrt{1-u^2}$, with $\tilde M_0$ given in (\ref{m0(u)}).

If we use the facts that $\sgn\varphi_a = -\sgn\varphi_b$ and $\tilde M_0(0)=1$,  we may write the result as
\begin{equation}
    \begin{split}
        L_1=&\frac{|\bq_a||\bq_b|}{(2\pi)^2}\Bigg\{\\
        &|\sin\varphi_a|\cos\varphi_b[1-\tilde M_0(\cos\varphi_a^\perp)] \\
        +&|\sin\varphi_b|\cos\varphi_a[1-\tilde M_0(\cos\varphi_b^\perp)]\\
        -&\cos\varphi_a\cos\varphi_b \sgn(\varphi_a\varphi_a^\perp) \tilde D_0(\cos\varphi_a^\perp)\\
        -&\cos\varphi_a\cos\varphi_b \sgn(\varphi_b\varphi_b^\perp)\tilde D_0(\cos\varphi_b^\perp)\Bigg\}.
    \end{split}
\end{equation}
Finally, we note that $\cos\varphi_{a(b)}^\perp = |\sin\varphi_{a(b)}|$, and $\sgn(\varphi_{a(b)}\varphi_{a(b)}^\perp) = -\sgn(\cos\varphi_{a(b)})$  (see the definition of $\varphi_{a(b)}^\perp$ after Eq. \ref{l1 abs}).   This allows us to write the final expression as,
\begin{equation}
    L_1 = \frac{|\bq_a||\bq_b|}{(2\pi)^2}\left[\cos\varphi_a W(\cos\varphi_b) + \cos\varphi_b W(\cos\varphi_a)\right],
\end{equation}
where
\begin{equation}
    W(\cos\varphi) = |\cos\varphi|\tilde D_0(|\sin\varphi|)-|\sin\varphi|(\tilde M_0(|\sin\varphi|)-1).
\end{equation}
It can be checked using (\ref{m0(u)}) and (\ref{d0(u)}) that
\begin{equation}
    W(u) = \frac{1}{\pi}\left(1+\frac{1-u^2}{2u} \log\frac{1+u}{1-u}\right).
\end{equation}

Finally, using $|\bq_a\times\bq_b| = |\bq_a||\bq_b||\sin(\varphi_a-\varphi_b)|$ we conclude that
\begin{equation}\begin{split}
    L(\bq_a,\bq_b)& = \frac{|\bq_a||\bq_b|}{(2\pi)^2}\Big(- |\sin(\varphi_a-\varphi_b)|\\
  &  +\cos\varphi_a W(\cos\varphi_b)+\cos\varphi_b W(\cos\varphi_a)\Big).
\end{split}\end{equation}
Eq. \ref{L analytic} in the main text then follows after we identify the (positive) internal angles $\phi_{a,b,c}$ opposite $\bq_{a,b,c}$ in the triangle formed by $\bq_a$, $\bq_b$ and $\bq_c=-\bq_a-\bq_b$ as:  $\phi_a = |\varphi_b|$, $\phi_b = |\varphi_a|$, $\phi_c = \pi - \phi_a - \phi_b = \pi -\varphi_a + \varphi_b$.

\section{Third order for $F_0^{\sigma\sigma'}$}
\label{Appendix B}

\subsection{Evaluation of Feynman Diagrams}
\label{appendix b1}
\begin{figure}
    \centering
    \includegraphics[width=0.5\linewidth]{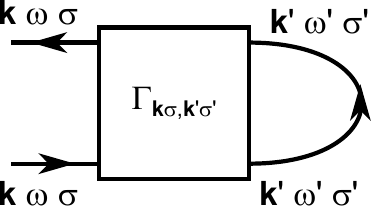}
    \caption{Vertex function as a part of a self-energy diagram.}
    \label{fig:self energy}
\end{figure}
 We evaluate the diagrams in Fig. \ref{Gamma diagrams} for vertex function $\Gamma^{\sigma\sigma'}_{\bk\bk'}$ using the standard Feynman rules in Euclidean time.   The diagrams are to be evaluated with the external legs removed, and with external Matsubara frequencies $i\omega$ and $i\omega'$ set to $E_F$.    To get the sign right it is useful to consider the vertex to be a part of a self-energy diagram, as shown in Fig. \ref{fig:self energy}.   Then, according to the Feynman rules, a diagram at $N$'th order in the interaction $u$ gets a factor  $(-u)^N (-1)^F$, where $F$ is the number of closed fermion loops.   Thus, the first order diagram in Fig. \ref{Gamma diagrams}a is simply $(-u)(-1)^{F=1} = u$.    Then, for example, the third order diagram in Fig. \ref{Gamma diagrams}e for $\Gamma^{\uparrow\uparrow}_{\bk\bk'}$ gives
 \begin{equation}\begin{split}
     2(-u)^3& (-1)^{F=1} \int \frac{d^2\bp d^2\bq}{(2\pi)^4} \int \frac{d\Omega d\nu}{(2\pi)^2} \\
    & G^\uparrow_{\bp,\Omega}G^\downarrow_{\bp+\bq,\Omega+\nu}G^\downarrow_{\bk+\bq,\omega+\nu}G^\downarrow_{\bk'+\bq,\omega'+\nu}
 \end{split}\end{equation}
 with
 \begin{equation}
     G^\sigma_{\bk,\omega} = \frac{1}{i\omega - E_\bk}.
 \end{equation}
The factor of $2$ accounts for a second similar diagram with the two interaction lines on the bottom instead of the top. 
Performing the frequency integrals by contour integration then leads to the expression for the contribution to $f_{\bk \bk'}^{\uparrow\uparrow}$ displayed in the second line of Eqs. \ref{fupup} plus a term with $\bk$ and $\bk'$ exchanged.    The other diagrams are evaluated similarly.

\subsection{Specification of Integrals}
\label{appendix b2}

We now provide some explanation for the evaluation of the integrals in Eqs. \ref{fupdown} and \ref{fupup}.   We begin by defining the dimensionless integrals $I_{1,2,3,4}$ specified in (\ref{F0I}).   

The first ${\cal O}(u^2)$ terms in (\ref{fupdown}) and (\ref{fupup}) are the same.  After substituting $E_\bk = |\bk|^2/(2m)$, $N_0=m/(2\pi)$ and using (\ref{F0 avarage}) we have
\begin{equation}
    I_1 = \left\langle\int\frac{d^2\bp }{2\pi} \frac{f_{\bp - \bk}-f_{\bp-\bk'}}{\bp \cdot (\bk -\bk')}\right\rangle_{\varphi,\varphi'},
    \label{i1 int}
\end{equation}
where as in (\ref{F0 avarage}) we specify $\bk = k_F \hat x$ and $\bk' = k_F(\cos\varphi \hat x + \sin\varphi \hat y).$
Similarly, the second ${\cal O}(u^2)$ term in (\ref{fupdown}) gives
\begin{equation}
    I_2 =  \left\langle \int \frac{d^2\bp }{2\pi} \frac{1-f_{\bp+\bk}-f_{\bp-\bk'}}{\bp\cdot(\bk'-\bk)-|\bp|^2}\right\rangle_{\varphi,\varphi'}.
        \label{i2 int}
\end{equation}

The ${\cal O}(u^3)$ correction to $\Gamma^{\uparrow\uparrow}_{\bk\bk'}$ involves two terms associated with the diagrams in Figs. \ref{Gamma diagrams}e,f, which are represented in the second and third lines of (\ref{fupup}).   There is an additional copy of each term with $\bk$ and $\bk'$ interchanged.   However, since we are computing the average over the directions of both $\bk$ and $\bk'$ those contributions are identical and just give a factor of two.   The resulting integrals are
\begin{equation}\begin{split}
    I_3 &=   \left\langle\int \frac{d^2\bp d^2\bq}{(2\pi)^2} \frac{4(f_\bq \bar f_{\bp-\bq}  f_{\bp-\bk}+\bar f_\bq f_{\bp-\bq} \bar f_{\bp-\bk})}{\bp\cdot(\bk'-\bk)
    \bp\cdot(\bk-\bq)}\right\rangle_{\varphi,\varphi'},
    \label{i3 int}
\end{split}\end{equation}
and
\begin{equation}\begin{split}
    I_4 &= \left\langle\int \frac{d^2\bp d^2\bq}{(2\pi)^2} \frac{4(f_\bq f_{\bp-\bq} \bar f_{\bp-\bk}+\bar f_\bq \bar f_{\bp-\bq} f_{\bp-\bk})}{\bp\cdot(\bk'-\bk)
    (\bk+\bq-\bp)\cdot(\bk-\bq)} \right\rangle_{\varphi,\varphi'}.
    \label{i4 int}
\end{split}\end{equation}
The integrals $I_2$ and $I_4$ are logarithmically divergent.   We regularize the divergence by restricting the ${\bf k}$ and ${\bf p}$ integrals to a large disk of radius $k_\Lambda$.    Since $I_{1,2,3,4}$ are dimensionless, we can set $k_F=1$.   $f_\bk$ is then non-zero inside the unit disk, and the divergent integrals are cut off by a disk of radius $k_\Lambda/k_F$.

\subsection{Evaluation of integrals}
\label{appendix b3}
The results for the evaluation of $I_{1,2,3,4}$ are reported in Eq. \ref{i results}.  We now provide some information to explain how we obtained them.   The calculations were long, so we will not provide all of the details, but we will try to make our path to the solution clear.

The following facts were helpful:

\begin{lemma}
For a real number $a$, and complex number $b$ with non-zero imaginary part, 
\begin{equation}
  C(a,b) =   \int_0^{2\pi} \frac{d\varphi}{2\pi}\frac{1}{a\cos\varphi + b} = \frac{\sgn {\rm Re}[b]}{\sqrt{b^2-a^2}}.
 \end{equation}
\begin{proof}
This is established by writing $\cos\varphi = (z + 1/z)/2$ and $d\varphi = -i dz/z$ and performing a contour integral of $z$ around the unit circle.
\end{proof} \label{lemma1}
\end{lemma}

\begin{lemma}
For real numbers $a$ and $b$
\begin{equation}
    \left\langle \frac{1}{a\cos\varphi+b}\right\rangle_\varphi = \frac{\sgn b}{\sqrt{b^2-a^2}}\theta(b^2-a^2).
\end{equation}
\begin{proof}
    The principle part of the integral in Lemma \ref{lemma1} is $\lim_{\epsilon\rightarrow 0} {\rm Re}[C(a,b+i\epsilon)]$.
\end{proof}
\label{lemma2}
\end{lemma}

\begin{lemma}
    \begin{equation}
        \left\langle \frac{1}{\cos\varphi - \cos\varphi'} \right\rangle_{\varphi'} = \frac{\pi}{2}( \delta(\varphi) - \delta(\varphi-\pi)).
    \end{equation}
    \begin{proof}
        The principle part in Lemma (\ref{lemma1}) is now $\lim_{\epsilon\rightarrow 0}{\rm Re}[\sgn(\cos\varphi)/\sqrt{i\epsilon - \sin^2\varphi}]$.  This is only non-zero in the vicinity of $\varphi=0,\pi$.  The coefficient of the $\delta$-function is determined by integrating over a neighborhood around those points.
    \end{proof}
    \label{lemma3}
\end{lemma}

\begin{lemma}
When $\bk$ and $\bk'$ are averaged over the unit circle
\begin{equation}
    \left\langle \frac{F(\bk)}{\bp\cdot(\bk - \bk')} \right\rangle_{\varphi,\varphi'} = \frac{F(\hat \bp) - F(-\hat\bp)}{4|\bp|}.
\end{equation}
\begin{proof}
    This follows directly from Lemma \ref{lemma3}.
\end{proof}
\label{lemma4}
\end{lemma}

We now proceed to evaluate the integrals.

\subsubsection{$I_1$}

Applying Lemma \ref{lemma4} to (\ref{i1 int}) after exchanging $\bk$ and $\bk'$ in the term involving $f_{\bp-\bk'}$ we obtain
\begin{equation}
    I_1 = \int \frac{d^2\bp}{2\pi}  \frac{2(f_{\bp -\hat \bp}-f_{\bp + \hat \bp})}{4|\bp|} = 1,
\end{equation}
where we used $f_{\bp -\hat\bp} = \theta(2-|\bp|)$ and $f_{\bp+\hat\bp}=0$.

\subsubsection{$I_2$}
$I_2$ can be evaluated by shifting $\bp$ in each of the terms (\ref{i2 int}) to make the argument of $f_\bp$ be $\bp$.   Then the integral over $\bp$ can be performed in polar coordinates.  The angular integral can be done with the aid of Lemma \ref{lemma2}.  After some simplification this leads to
\begin{equation}
    \begin{split}
        I_2 = \Bigg\langle&\int_0^{k_\Lambda}
        \frac{p dp}{k_F^2\sin^2\frac{1}{2}\theta_{\bk \bk'} - p^2} \\
        + &\int_0^{k_F\sqrt{1-|\sin\theta_{\bk\bk'}|}}\frac{2 p dp}{\sqrt{(p^2-k_F^2)^2-k_F^4\sin^2\theta_{\bk\bk'}}}
        \Bigg]\Bigg\rangle_{\varphi,\varphi'},
    \end{split}
\end{equation}
where $\theta_{\bk\bk'} = \varphi-\varphi'$.   Performing the integral over $p$ followed by the angular averages then gives
\begin{equation}\begin{split}
    I_2 &= - \left\langle \log\left(\frac{k_\Lambda}{k_F}\frac{1}{|\cos\frac{1}{2}\theta_{\bk\bk'}|}\right)\right\rangle_{\varphi,\varphi'}\\
    &=  - \log \frac{k_\Lambda}{k_F}- \log 2.
\end{split}\end{equation}

\subsubsection{$I_3$}
We first use Lemma \ref{lemma4} to perform the angular averages in Eq. \ref{i3 int}, again noting that $f_{\bp-\hat\bp}=\theta(2-|\bp|)$ and $f_{\bp+\hat\bp}=0$.   In addition, we write $\bar f_{\bp+\hat \bp} =1 = \theta(2-|\bp|)+\theta(|\bp|-2)$.
\begin{equation}\begin{split}
    I_3 = &\int \frac{d^2\bp d^2\bq}{(2\pi)^2} \frac{\theta(2-|\bp|)}{|\bp|}
 \left[\frac{f_\bq \bar f_{\bp-\bq}}{|\bp|-\bp\cdot\bq} + \frac{\bar f_\bq f_{\bp-\bq}}{|\bp|+\bp\cdot\bq}\right] \\
+&\int \frac{d^2\bp d^2\bq}{(2\pi)^2} \frac{\theta(|\bp|-2)}{|\bp|}
\left[\frac{\bar f_\bq f_{\bp-\bq}}{|\bp|-\bp\cdot\bq} + \frac{\bar f_\bq f_{\bp-\bq}}{|\bp|+\bp\cdot\bq}\right].
\label{i3 v1}
\end{split}
\end{equation}
The remaining integral over $\bp$ and $\bq$ is invariant under rotations, so we can do one of the angular integrals and set $\bp = 2 t \hat x$.     
We found it useful to shift 
$\bq \rightarrow \bs + \bp/2 = \bs + t \hat x$, which then gives
\begin{equation}\begin{split}
    I_3   = &\int_0^1 dt \int \frac{d^2\bs}{2\pi} \left[\frac{\bar f_{\bs + t\hat x}f_{\bs-t \hat x}}{t(t+1)+ t s_x}
    -\frac{f_{\bs+t\hat x}\bar f_{\bs-t\hat x}}{t(t-1)+t s_x}\right] \\
  +&  \int_1^\infty dt \int \frac{d^2\bs}{2\pi} \left[\frac{\bar f_{\bs + t\hat x}f_{\bs-t \hat x}}{t(t+1)+ t s_x}
    -\frac{\bar f_{\bs+t\hat x}f_{\bs-t\hat x}}{t(t-1)+t s_x}\right] .
    \label{i3 v2}
\end{split}\end{equation}
It was most convenient to perform the integral over $\bs$ in Cartesion coordinates, performing the $s_y$ integral first, which after some simplification gives
\begin{equation}
    \begin{split}
        = \int_0^1 \frac{dt}{\pi t} \Bigg[  &\int_{-t}^{1}dx \sqrt{1-x^2}\left(\frac{1}{x+2t+1}+\frac{1}{x+1}\right)\\
      -  & \int_{t}^{1}dx \sqrt{1-x^2} \left(\frac{1}{x+1}+\frac{1}{x+1-2t}\right)\Bigg] \\
      +\int_1^\infty \frac{dt}{\pi t}&\int_{-1}^{1}dx \sqrt{1-x^2}\left(\frac{1}{x+2t+1+x}-\frac{1}{x+2t-1}\right).
    \end{split}
\end{equation}

The integral over $x$ ($=s_x$) can be done by applying (\ref{fintegral}), which after some simplification leads to
\begin{equation}
    \begin{split}
        =\int_0^1 \frac{2 dt}{\pi t}\Bigg[& \pi t + 2\sin^{-1} t-\sqrt{t(t+1)}\cos^{-1}(1-2t)\\
      &  -\sqrt{t(1-t)}\cosh^{-1}(1+2t)
        \Bigg]\\
        +\int_1^\infty \frac{2dt}{t} &\left[ 1 - \sqrt{t(t+1)}+\sqrt{t(t-1)}\right].
    \end{split}
\end{equation}

Performing the final integral over $t$ then gives
\begin{equation}
    I_3 = 2(1-\log 2).
    \label{i3 result}
\end{equation}
Eq. \ref{i3 result} can be verified by numerically integrating (\ref{i3 v2}).

\subsubsection{$I_4$}
We apply the same steps as in (\ref{i3 v1}) to (\ref{i4 int}), also noting that
the term involving $f_\bq f_{\bp-\bq} \bar f_{\bp-\hat \bp}$ vanishes because $f_\bq f_{\bp-\bq}$ is only nonzero for $|\bp|<2$, while $\bar f_{\bp-\hat \bp} = \theta(|\bp|-2)$.   This then gives
\begin{equation}\begin{split}
    I_4 &= \int \frac{d^2\bp d^2\bq}{(2\pi)^2} \frac{\theta(2-|\bp|)}{|\bp|} \\
 &  \left[\frac{\bar f_\bq \bar f_{\bp-\bq}}{1-|\bp|-\bq\cdot(\bq-\bp)} 
 -\frac{f_\bq f_{\bp-\bq}}{1+|\bp|-\bq\cdot(\bq-\bp)}\right].
 \label{i4 v1}
\end{split}\end{equation}

As in (\ref{i3 v2}) after doing the angular integral we set $\bp = 2 t \hat x$ and
$\bq \rightarrow \bs + \bp/2 = \bs + t \hat x$, which gives
\begin{equation}
    = \int_0^1 dt \int \frac{d^2\bs}{\pi} \left[\frac{\bar f_{\bs + t\hat x}\bar f_{\bs-t \hat x}}{(1-t)^2 - |\bs|^2}
    -\frac{f_{\bs+t\hat x}f_{\bs-t\hat x}}{(1+t)^2-|\bs|^2}\right].
    \label{i4 v2}
\end{equation}
We did the integral over $\bs$ using polar coordinates $\bs = s(\cos\varphi,\sin\varphi)$, performing the radial integral first.   The limits on the radial integral are $k_\Lambda/k_F$ (for the first term), along with
the points obtained by solving $s^2+t^2\pm 2 s t \cos\varphi = 1$ for $s$.   After some simplification this gave,
\begin{equation}\begin{split}
    = -\frac{4}{\pi} &\int_0^1 dt \int_0^{\pi/2} d\varphi \Big\{\log\left[\frac{k_\Lambda}{k_F} \frac{1+t}{2t}\right]\\
  &  -\log(\cos\varphi + \sqrt{1-t^2 \sin^2\varphi})\Big\}.
\end{split}\end{equation}
The integrals in the first log are elementary.   For the second log, we use
\begin{equation}
    \int_0^1 dt \log(\sqrt{1-t^2\sin^2\varphi}+\cos\varphi) = -1 + \log 2 + \varphi \cot \varphi
\end{equation}
which then allows us to conclude that
\begin{equation}
    I_4 = -2 \log\frac{k_\Lambda}{k_F} - 2(1-\log 2).
    \label{i4 result}
\end{equation}
Eq. \ref{i4 result} can be verified by numerically integrating (\ref{i4 v2}) after subtracting $\theta(|\bs|-1)/(\pi|\bs|^2)$ to cancel the log divergence.

\end{document}